\documentclass[aps,prb,reprint,showpacs,superscriptaddress,preprintnumbers]{revtex4-1}
\usepackage{amsmath,txfonts,bm,graphicx,color,hyperref}
\allowdisplaybreaks[4]
\DeclareMathOperator\tr{tr}
\begin{document}
\title{Orbital angular momentum in a topological superconductor with Chern number higher than $1$}
\author{Atsuo Shitade}
\affiliation{Yukawa Institute for Theoretical Physics, Kyoto University, Kyoto 606-8502, Japan}
\altaffiliation[Present address: ]{RIKEN Center for Emergent Matter Science, 2-1 Hirosawa, Wako, Saitama 351-0198, Japan}
\author{Yuki Nagai}
\affiliation{CCSE, Japan Atomic Energy Agency, 178-4-4, Wakashiba, Kashiwa, Chiba, 277-0871, Japan}
\date{\today}
\begin{abstract}
  We investigate the bulk orbital angular momentum (AM) in a two-dimensional hole-doped topological superconductor (SC)
  which is composed of a hole-doped semiconductor thin film, a magnetic insulator, and an $s$-wave SC and is
  characterized by the Chern number $C = -3$.
  In the topological phase, $L_z/N$ is strongly reduced from the intrinsic value by the non-particle-hole-symmetric edge states
  as in the corresponding chiral $f$-wave SCs when the spin-orbit interactions (SOIs) are small,
  while this reduction of $L_z/N$ does not work for the large SOIs.
  Here $L_z$ and $N$ are the bulk orbital AM and the total number of particles at zero temperature, respectively.
  As a result, $L_z/N$ is discontinuous or continuous at the topological phase transition depending on the strengths of the SOIs.
  We also discuss the effects of the edge states by calculating the radial distributions of the orbital AM.
\end{abstract}
\maketitle
\section{Introduction} \label{sec:intro}
Orbital angular momentum (AM) is one of the most fundamental quantities in classical and quantum mechanics.
In condensed matter physics, the bulk orbital AM in chiral superconductors (SCs) has been intensively studied~\cite{%
Ishikawa01061977,PhysRevLett.43.596,PhysRevB.21.980,Ishikawa01041980,Volovik1995,JPSJ.67.216,Goryo1998549,%
PhysRevB.69.184511,PhysRevLett.101.150409,PhysRevA.81.053605,PhysRevB.84.214509,PhysRevB.85.100506,PhysRevB.90.134510,%
PhysRev.123.1911,RevModPhys.47.331,Volovik1975,Cross1975,PhysRevLett.114.195301,Volovik2014,PhysRevB.90.224519}.
A chiral SC is an exotic SC whose Cooper pairs carry nonzero orbital AM $\ell$, and hence the time-reversal symmetry is spontaneously broken.
$^3$He-A is the only material experimentally established as a chiral superfluid, whose pairing symmetry is chiral $p$ wave~\cite{RevModPhys.47.331}.
Also, Sr$_2$RuO$_4$ is widely believed to be a chiral $p$-wave SC~\cite{RevModPhys.75.657,JPSJ.81.011009}.
Therefore, most of the previous studies have focused on chiral $p$ wave
and at least theoretically concluded $L_z/N = +1/2$ in an ideal situation,
where $L_z$ and $N$ are the bulk orbital AM and the total number of particles at zero temperature, respectively~\cite{%
Ishikawa01061977,PhysRevLett.43.596,PhysRevB.21.980,Ishikawa01041980,Volovik1995,JPSJ.67.216,Goryo1998549,%
PhysRevB.69.184511,PhysRevLett.101.150409,PhysRevA.81.053605,PhysRevB.84.214509,PhysRevB.85.100506,PhysRevB.90.134510}.
In general, chiral SCs with $\ell$ are expected to have $L_z/N = \ell/2$ without an edge or a vortex~\cite{PhysRevB.90.134510}, which we call the intrinsic value.

Recently, it was found that the bulk orbital AM in two-dimensional chiral SCs with $\ell > 1$
is remarkably reduced as $L_z/N = o(1)$~\cite{PhysRevLett.114.195301,Volovik2014,PhysRevB.90.224519}.
In two dimensions, chiral SCs are gapped and are topological SCs characterized by the Chern number $C = \ell$ and the presence of the Majorana edge states.
When $\ell = 1$, only one edge state is particle-hole symmetric (PHS) by itself and is called the PHS edge state~\cite{PhysRevLett.114.195301}.
On the other hand, when $\ell > 1$ is even, each edge state is PHS with another but not PHS by itself.
Such edge states are called the non-PHS edge states~\cite{PhysRevLett.114.195301}.
When $\ell > 1$ is odd, there are the PHS and non-PHS edge states.
The non-PHS edge states give rise to the nonzero spectral asymmetry and reduce the bulk orbital AM from the intrinsic value $L_z/N = \ell/2$.
Although chiral $d$ or $f$ wave has been proposed in UPt$_3$~\cite{RevModPhys.74.235},
URu$_2$Si$_2$~\cite{PhysRevLett.99.116402,PhysRevLett.100.017004,1367-2630-11-5-055061,JPSJ.81.023704},
and SrPtAs~\cite{PhysRevB.86.100507,PhysRevB.87.180503,PhysRevB.89.020509},
the theory cannot be applied to these three-dimensional materials with nodes directly.

Apparently, the above reduction of the bulk orbital AM requires not $\ell > 1$ but $C > 1$.
Realization of two-dimensional time-reversal broken topological SCs with $C > 1$ is classified into two types.
The first type is intrinsic, namely, without impurities, and can be realized in a quantum anomalous Hall insulator in proximity to an $s$-wave SC~\cite{PhysRevB.82.184516},
a heterostructure of a hole-doped semiconductor thin film, a magnetic insulator, and an $s$-wave SC~\cite{PhysRevLett.106.157003},
and the systems with the help of $p$-wave SCs~\cite{PhysRevB.79.094504,PhysRevB.88.184513,PhysRevB.90.064507}.
The second type relies on impurities and can be realized
in a lattice of magnetic impurities on the surface of an $s$-wave SC with the Rashba spin-orbit interaction (SOI)~\cite{PhysRevLett.114.236803,1501.00999}
and that of nonmagnetic impurities in a chiral $p$-wave SC~\cite{PhysRevB.93.035134}.
In these proposals, arbitrarily high Chern numbers are available in principle.
In return for using a conventional $s$-wave SC, SOIs play an important role in generating the effective chirality of Cooper pairs.
Among these proposals, a hole-doped topological SC proposed in Ref.~\onlinecite{PhysRevLett.106.157003} is a natural extension
of an electron-doped topological SC composed of an electron-doped semiconductor thin film, a magnetic insulator, and an $s$-wave SC~\cite{PhysRevLett.101.160401,PhysRevLett.103.020401,PhysRevLett.104.040502}
and offers an experimentally feasible system to verify the theory of the bulk orbital AM.
We note that hole-doped semiconductors accompanied by ferromagnetism can be realized by Mn doping
and are called dilute magnetic semiconductors~\cite{RevModPhys.78.809,RevModPhys.86.187}.
However, it is not trivial how the bulk orbital AM is reduced in this system because the orbital AM is not conserved in the presence of the SOIs.

In this paper, we investigate the bulk orbital AM in a hole-doped topological SC~\cite{PhysRevLett.106.157003}
comparing with that in an electron-doped topological SC~\cite{PhysRevLett.101.160401,PhysRevLett.103.020401,PhysRevLett.104.040502}.
We calculate the bulk orbital AM both by the Green's function formula~\cite{PhysRevB.90.134510} and on a circular disk.
As mentioned above, the non-PHS edge states dramatically reduce the bulk orbital AM in the case of $C > 1$.
In addition, Tsutumi and Machida pointed out that $L_z/N = +1/2$ in chiral $p$-wave SCs
consists of $L_z^{\rm MJ}/N = +1$ from the Majorana edge state and $L_z^{\rm cont} = -1/2$ from the continuum states~\cite{PhysRevB.85.100506}.
The authors already found that $L_z/N$ in an electron-doped topological SC is continuous at the topological phase transition
and is nonzero even in the trivial phase which does not support the Majorana edge state~\cite{PhysRevB.92.024502}.
Therefore it is important to reveal the difference of the contributions to the bulk orbital AM from the PHS edge, non-PHS edge, and continuum states.
We also discuss the effects of the SOIs which are indispensable for topological SCs.

We find that the conserved quantity is modified by SOIs.
When the Chern number is higher than $1$, it can be nonzero owing to the presence of the non-PHS edge states and the nonzero spectral asymmetry,
which results in the reduction of the bulk orbital AM as shown in Refs.~\onlinecite{PhysRevLett.114.195301,Volovik2014,PhysRevB.90.224519}.
Since the bulk orbital AM calculated in the reciprocal space is continuous at the topological phase transition, that calculated in the real space shows a jump.
Differently from the existing literature, this reduction is not universal and can be tuned by SOIs in this system.

\section{Model} \label{sec:model}
First, we review a hole-doped topological SC composed of a hole-doped semiconductor thin film, a magnetic insulator, and an $s$-wave SC~\cite{PhysRevLett.106.157003},
\begin{align}
  {\cal H}_{\vec k}
  = & [\xi_{1k} - \sqrt{3} \xi_{2k} (\Gamma_4 \cos 2 \phi + \Gamma_3 \sin 2 \phi) \notag \\
  & + (\xi_{2k} - W) \Gamma_5 + 2 \alpha k (j_y \cos \phi - j_x \sin \phi)] \tau_z - 2 h j_z \notag \\
   & + [(\Delta_{\rm H} - \Delta_{\rm L})/2 + (\Delta_{\rm H} + \Delta_{\rm L}) \Gamma_5/2] \tau_x, \label{eq:luttingertsc-hk}
\end{align}
where we choose the Nambu basis, $\Psi_{\vec k} = [c_{{\vec k} +3/2}, c_{{\vec k} +1/2}, c_{{\vec k} -1/2}, c_{{\vec k} -3/2},
c_{-{\vec k} -3/2}^{\dag}, -c_{-{\vec k} -1/2}^{\dag}, c_{-{\vec k} +1/2}^{\dag}, -c_{-{\vec k} +3/2}^{\dag}]^{\rm T}$.
${\vec \tau}$ is a set of the Pauli matrices for the Nambu space,
and ${\vec j}$ is the total AM of $j = 3/2$ holes but is called spin to avoid confusion below.
We use the standard representation,
\begin{subequations} \begin{align}
  j_x
  = &
  \begin{bmatrix}
    & \sqrt{3}/2 & & \\
    \sqrt{3}/2 & & 1 & \\
    & 1 & & \sqrt{3}/2 \\
    & & \sqrt{3}/2 &
  \end{bmatrix}, \label{eq:j3/2_x} \\
  j_y
  = &
  \begin{bmatrix}
    & -\sqrt{3} i/2 & & \\
    \sqrt{3} i/2 & & -i & \\
    & i & & -\sqrt{3} i/2 \\
    & & \sqrt{3} i/2 &
  \end{bmatrix}, \label{eq:j3/2_y} \\
  j_z
  = &
  \begin{bmatrix}
    3/2 & & & \\
    & 1/2 & & \\
    & & -1/2 & \\
    & & & -3/2
  \end{bmatrix}, \label{eq:j3/2_z}
\end{align} \label{eq:j3/2} \end{subequations}
and construct the $\Gamma$ matrices satisfying the anticommutation relations $\{\Gamma_a, \Gamma_b\}= 2 \delta_{ab}$~\cite{PhysRevB.69.235206},
\begin{subequations} \begin{align}
  \Gamma_1
  = & \{j_y, j_z\}/\sqrt{3}
  =
  \begin{bmatrix}
    & -i & & \\
    i & & & \\
    & & & i \\
    & & -i & 
  \end{bmatrix}, \label{eq:gamma3/2_1} \\
  \Gamma_2
  = & \{j_z, j_x\}/\sqrt{3}
  =
  \begin{bmatrix}
    & 1 & & \\
    1 & & & \\
    & & & -1 \\
    & & -1 & 
  \end{bmatrix}, \label{eq:gamma3/2_2} \\
  \Gamma_3
  = & \{j_x, j_y\}/\sqrt{3}
  =
  \begin{bmatrix}
    & & -i & \\
    & & & -i \\
    i & & & \\
    & i & &
  \end{bmatrix}, \label{eq:gamma3/2_3} \\
  \Gamma_4
  = & (j_x^2 - j_y^2)/\sqrt{3}
  =
  \begin{bmatrix}
    & & 1 & \\
    & & & 1 \\
    1 & & & \\
    & 1 & &
  \end{bmatrix}, \label{eq:gamma3/2_4} \\
  \Gamma_5
  = & j_z^2 - 5/4
  =
  \begin{bmatrix}
    1 & & & \\
    & -1 & & \\
    & & -1 & \\
    & & & 1
  \end{bmatrix}. \label{eq:gamma3/2_5}
\end{align} \label{eq:gamma3/2} \end{subequations}
The first, second, and third terms in Eq.~\eqref{eq:luttingertsc-hk} are the two-dimensional Luttinger Hamiltonian,
which describes a hole-doped semiconductor thin film~\cite{PhysRev.102.1030}.
We define $\xi_{1k} = \gamma_1 k^2/2 m - \mu$ and $\xi_{2k} = \gamma_2 k^2/2 m$,
in which ${\vec k} = k [\cos \phi, \sin \phi]$ is a two-dimensional wavevector,
$\gamma_1$ and $\gamma_2$ are the Luttinger parameters, and $\mu$ is the chemical potential.
In a hole-doped semiconductor, confinement into a quantum well yields quantization of momentum $\langle k_z \rangle = 0$ and $\langle k_z^2 \rangle \not= 0$
and opens up the gap $W = \gamma_2 \langle k_z^2 \rangle/m$ between the heavy-hole bands with $j_z = \pm 3/2$ and the light-hole bands with $j_z = \pm 1/2$.
This point is a crucial difference from an electron-doped semiconductor, in which confinement just shifts the chemical potential.
The fourth and fifth terms are the Rashba SOI and the Zeeman interaction, respectively,
and the sixth and seventh terms are the $s$-wave pairing potentials induced by the proximity effect.
Note that we neglect some terms allowed by the theory of invariants~\cite{PhysRev.102.1030}.

\begin{figure*}
  \centering
  \includegraphics[clip,width=0.67\textwidth]{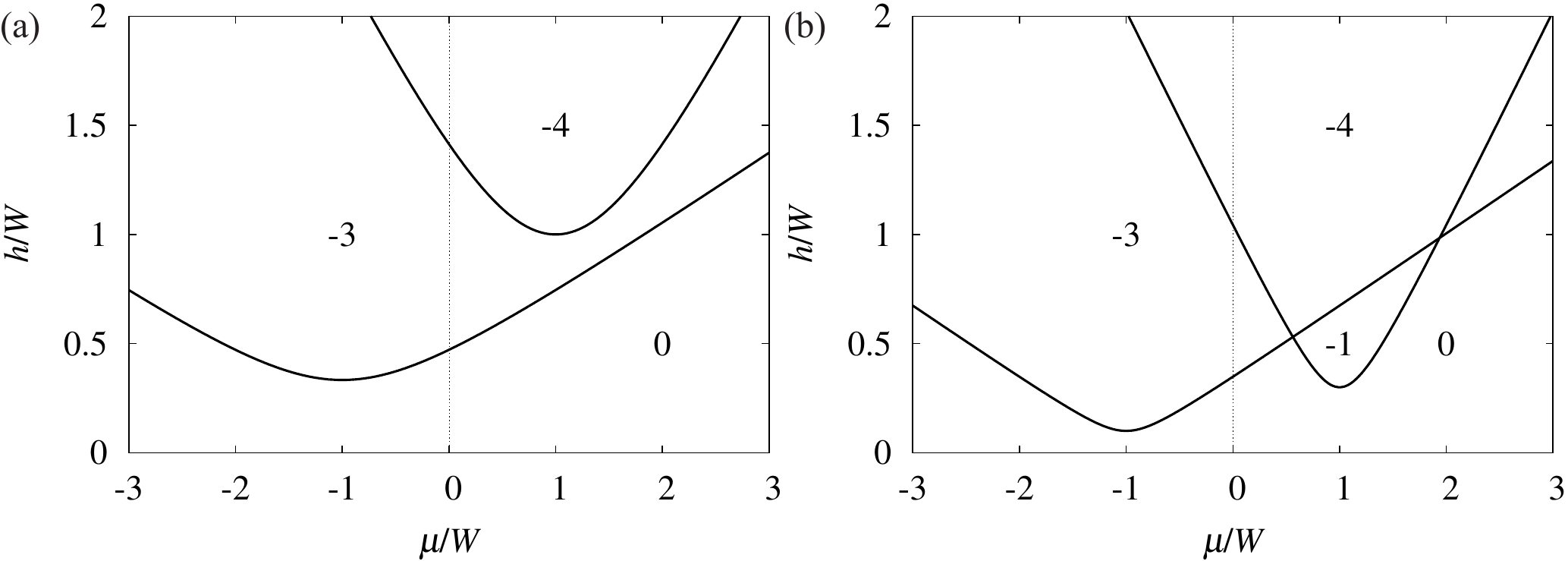}
  \caption{%
  Phase diagrams in terms of the Chern number
  for (a) $|\Delta_{\rm H}/W| = |\Delta_{\rm L}/W| = 1$
  and (b) $|\Delta_{\rm H}/W| = |\Delta_{\rm L}/W| = 0.35$.%
  } \label{fig:luttingertsc-phase}
\end{figure*}

The topological phase transitions occur when the gap closes at $k = 0$.
We can analytically obtain eight eigenvalues, $\pm 3 h \pm \sqrt{(\mu + W)^2 + \Delta_{\rm H}^2}$ and $\pm h \pm \sqrt{(\mu - W)^2 + \Delta_{\rm L}^2}$,
and two phase transition lines, $3 h_{\rm cH} = \sqrt{(\mu + W)^2 + \Delta_{\rm H}^2}$ and $h_{\rm cL} = \sqrt{(\mu - W)^2 + \Delta_{\rm L}^2}$,
where the Chern number changes by $-3$ and $-1$, respectively.
The phase diagrams for different $\Delta_{\rm H}$ and $\Delta_{\rm L}$ are shown in Fig.~\ref{fig:luttingertsc-phase}.
$C = -3$ is realized when the chemical potential lies in the lowest heavy-hole band alone,
which is described by the cubic Rashba model with the Zeeman interaction in the large-$W$ limit~\cite{PhysRevLett.106.157003}.
We should note that the gap may accidentally close at $k \not= 0$, which does not change the Chern number.

Below we put $\gamma_1 = 1$, $\gamma_2 = 0.3$, $m = 0.5$, $\mu = 0$, $W = 1$, $\Delta_{\rm H} = -\Delta_{\rm L} = 1$ focusing on Fig.~\ref{fig:luttingertsc-phase}(a),
leading to $h_{\rm cH} = 0.47$ and $h_{\rm cL} = 1.41$.
Once we choose energy and wavenumber units,
the remaining important parameters are $\gamma_2/\gamma_1$, $\mu/W$, $2 m \alpha^2/\gamma_1 W$, $h/W$, $\Delta_{\rm H}/W$, and $\Delta_{\rm L}/W$.
Among these, $\Delta_{\rm H}/W$ and $\Delta_{\rm L}/W$ drastically change the phase diagram as shown in Fig.~\ref{fig:luttingertsc-phase}.
Here we arbitrarily choose $\Delta_{\rm H}/W = -\Delta_{\rm L}/W = 1$ since we cannot estimate their experimental values.
$\gamma_2/\gamma_1$ takes $0.2$-$0.4$ for typical III-V semiconductors such as GaAs~\cite{9783540011873}, and hence our choice of $\gamma_2/\gamma_1 = 0.3$ is realistic.
$\mu/W$ can be tuned experimentally by doping carriers or applying an external field.
Note that $W$ is about $16 {\rm meV}$ for a GaAs quantum well with $100 {\rm \AA}$ thickness.

\section{Methods} \label{sec:methods}
We calculate the bulk orbital AM in two ways.
One is based on the Green's function formula in the reciprocal space~\cite{PhysRevB.90.134510}.
Since there is no edge state in the periodic boundary condition,
we do not discuss the effects of the non-PHS edge states but interpret the resulting bulk orbital AM as the intrinsic value.
Here we calculate the Hall viscosity $2 \eta_{\rm H}$,
which is known to be equal to the bulk orbital AM $L_z$ in two-dimensional gapped systems at zero temperature~\cite{PhysRevB.79.045308,PhysRevB.84.085316,PhysRevB.86.245309}.
By using the Matsubara Green's function ${\cal G}(i \omega_n, {\vec k}) = (i \omega_n - {\cal H}_{\vec k})^{-1}$,
the Chern number~\cite{9780199564842} and the Hall viscosity~\cite{PhysRevB.90.134510} can be evaluated by
\begin{subequations} \begin{align}
  C
  = & \frac{1}{6 i} \epsilon_{abc} \int \frac{{\rm d}^3 k}{(2 \pi)^2}
  \tr {\cal G} \partial_{k_a} {\cal G}^{-1} {\cal G} \partial_{k_b} {\cal G}^{-1} {\cal G} \partial_{k_c} {\cal G}^{-1}, \label{eq:luttingertsc-ck} \\
  2 \eta_{\rm H}
  = & \frac{1}{12 i} \epsilon_{abc} \int \frac{{\rm d}^3 k}{(2 \pi)^3} {\vec k}^2
  \tr {\cal G} \partial_{k_a} {\cal G}^{-1} {\cal G} \partial_{k_b} {\cal G}^{-1} {\cal G} \partial_{k_c} {\cal G}^{-1}. \label{eq:luttingertsc-lzk}
\end{align} \label{eq:luttingertsc-k} \end{subequations}
Here we introduce $k_0 = i \omega_n$, and $\epsilon_{abc}$ is the totally antisymmetric tensor with $\epsilon_{012} = 1$.
Although the integrals are carried out over the $(i \omega_n, {\vec k})$ space, the prefactor ${\vec k}^2$ in Eq.~\eqref{eq:luttingertsc-lzk} is in two dimensions.

\begin{figure*}
  \centering
  \includegraphics[clip,width=0.99\textwidth]{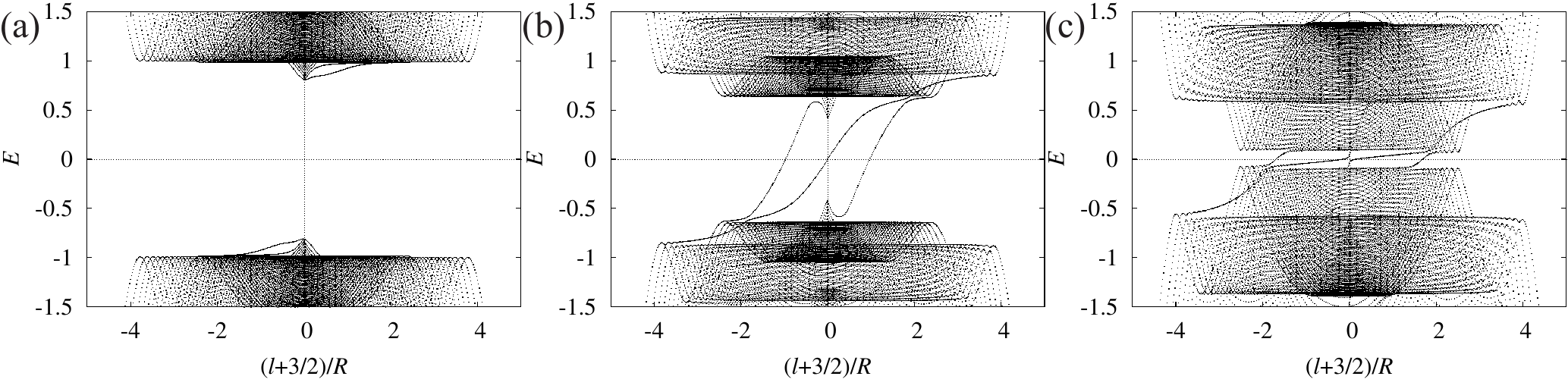}
  \caption{%
  Spectra
  for (a) $h = 0.2 < h_{\rm cH}$ ($C = 0$),
  (b) $h_{\rm cH} < h = 1 < h_{\rm cL}$ ($C = -3$),
  and (c) $h = 2 > h_{\rm cL}$ ($C = -4$).
  We put $\alpha = 1.5$.%
  } \label{fig:luttingertsc-eigen}
\end{figure*}

\begin{widetext}
The other way is a straightforward calculation in the real space.
We solve the Bogoliubov-de Gennes (BdG) equation ${\cal H}_{-i {\vec \nabla}} \Psi_{\ell i}({\vec r}) = E_{\ell i} \Psi_{\ell i}({\vec r})$ on a disk with its radius $R$.
Since the system preserves the rotational symmetry, the BdG wave function can be expressed as
\begin{equation}
  \Psi_{\ell i}({\vec r})
  = (2 \pi)^{-1/2} e^{i \ell \phi} [u_{\ell i 0}(r), e^{i \phi} u_{\ell i 1}(r), e^{2 i \phi} u_{\ell i 2}(r), e^{3 i \phi} u_{\ell i 3}(r),
  v_{\ell i 0}(r), e^{i \phi} v_{\ell i 1}(r), e^{2 i \phi} v_{\ell i 2}(r), e^{3 i \phi} v_{\ell i 3}(r)]^{\rm T}, \label{eq:luttingertsc-wfr}
\end{equation}
and the radial BdG equation is written by
\begin{subequations} \begin{align}
  & [-(\gamma_1 + \gamma_2) {\cal L}_{\ell}(r)/2m - \mu - W - 3 h] u_{\ell i 0}(r) - \sqrt{3} \alpha {\cal R}_{\ell + 1}(r) u_{\ell i 1}(r)
  + (\sqrt{3} \gamma_2/2m) {\cal S}_{\ell + 2}(r) u_{\ell i 2}(r) \notag \\
  & + \Delta_{\rm H} v_{\ell i 0}(r)
  = E_{\ell i} u_{\ell i 0}(r), \label{eq:luttingertsc-bdg0} \\
  & \sqrt{3} \alpha {\cal R}_{-\ell}(r) u_{\ell i 0}(r) + [-(\gamma_1 - \gamma_2) {\cal L}_{\ell + 1}(r)/2m - \mu + W - h] u_{\ell i 1}(r)
  - 2 \alpha {\cal R}_{\ell + 2}(r) u_{\ell i 2}(r) + (\sqrt{3} \gamma_2/2m) {\cal S}_{\ell + 3}(r) u_{\ell i 3}(r) \notag \\
  & - \Delta_{\rm L} v_{\ell i 1}(r)
  = E_{\ell i} u_{\ell i 1}(r), \label{eq:luttingertsc-bdg1} \\
  & (\sqrt{3} \gamma_2/2m) {\cal S}_{-\ell}(r) u_{\ell i 0}(r) + 2 \alpha {\cal R}_{-(\ell + 1)}(r) u_{\ell i 1}(r)
  + [-(\gamma_1 - \gamma_2) {\cal L}_{\ell + 2}(r)/2m - \mu + W + h] u_{\ell i 2}(r) - \sqrt{3} \alpha {\cal R}_{\ell + 3}(r) u_{\ell i 3}(r) \notag \\
  & - \Delta_{\rm L} v_{\ell i 2}(r)
  = E_{\ell i} u_{\ell i 2}(r), \label{eq:luttingertsc-bdg2} \\
  & (\sqrt{3} \gamma_2/2m) {\cal S}_{-(\ell + 1)}(r) u_{\ell i 1}(r) + \sqrt{3} \alpha {\cal R}_{-(\ell + 2)}(r) u_{\ell i 2}(r)
  + [-(\gamma_1 + \gamma_2) {\cal L}_{\ell + 3}(r)/2m - \mu - W + 3 h] u_{\ell i 3}(r) \notag \\
  & + \Delta_{\rm H} v_{\ell i 3}(r)
  = E_{\ell i} u_{\ell i 3}(r), \label{eq:luttingertsc-bdg3} \\
  & -[-(\gamma_1 + \gamma_2) {\cal L}_{\ell}(r)/2m - \mu - W + 3 h] v_{\ell i 0}(r) + \sqrt{3} \alpha {\cal R}_{\ell + 1}(r) v_{\ell i 1}(r)
  - (\sqrt{3} \gamma_2/2m) {\cal S}_{\ell + 2}(r) u_{\ell i 2}(r) \notag \\
  & + \Delta_{\rm H} u_{\ell i 0}(r)
  = E_{\ell i} v_{\ell i 0}(r), \label{eq:luttingertsc-bdg4} \\
  & -\sqrt{3} \alpha {\cal R}_{-\ell}(r) v_{\ell i 0}(r) - [-(\gamma_1 - \gamma_2) {\cal L}_{\ell + 1}(r)/2m - \mu + W + h] v_{\ell i 1}(r)
  + 2 \alpha {\cal R}_{\ell + 2}(r) v_{\ell i 2}(r) - (\sqrt{3} \gamma_2/2m) {\cal S}_{\ell + 3}(r) v_{\ell i 3}(r) \notag \\
  & - \Delta_{\rm L} u_{\ell i 1}(r)
  = E_{\ell i} v_{\ell i 1}(r), \label{eq:luttingertsc-bdg5} \\
  & -(\sqrt{3} \gamma_2/2m) {\cal S}_{-\ell}(r) v_{\ell i 0}(r) - 2 \alpha {\cal R}_{-(\ell + 1)}(r) v_{\ell i 1}(r)
  - [-(\gamma_1 - \gamma_2) {\cal L}_{\ell + 2}(r)/2m - \mu + W - h] v_{\ell i 2}(r) + \sqrt{3} \alpha {\cal R}_{\ell + 3}(r) v_{\ell i 3}(r) \notag \\
  & - \Delta_{\rm L} u_{\ell i 2}(r)
  = E_{\ell i} v_{\ell i 2}(r), \label{eq:luttingertsc-bdg6} \\
  & -(\sqrt{3} \gamma_2/2m) {\cal S}_{-(\ell + 1)}(r) v_{\ell i 1}(r) - \sqrt{3} \alpha {\cal R}_{-(\ell + 2)}(r) v_{\ell i 2}(r)
  - [-(\gamma_1 + \gamma_2) {\cal L}_{\ell + 3}(r)/2m - \mu - W - 3 h] v_{\ell i 3}(r) \notag \\
  & + \Delta_{\rm H} u_{\ell i 3}(r)
  = E_{\ell i} v_{\ell i 3}(r), \label{eq:luttingertsc-bdg7}
\end{align} \label{eq:luttingertsc-bdg} \end{subequations}
Here we introduce ${\cal L}_{\ell}(r) \equiv \partial_r^2 + (1/r) \partial_r - \ell^2/r^2$, ${\cal R}_{\ell + 1}(r) \equiv \partial_r + (\ell + 1)/r$,
and ${\cal S}_{\ell + 2}(r) \equiv \partial_r^2 + [(2 \ell + 3)/r] \partial_r + \ell (\ell + 2)/r^2$.
We impose the boundary conditions $g^{\prime}(0) = g(R) = 0$ ($g = u_{\ell i m}, v_{\ell i m}$) and the normalization condition,
\begin{equation}
  \int_0^R r {\rm d} r [{\bm u}_{\ell i}^2(r) + {\bm v}_{\ell i}^2(r)]
  = 1. \label{eq:rashbatsc-norm}
\end{equation}
Owing to the PHS ${\cal H}_{\vec k} = -C {\cal H}^{\ast}_{-{\vec k}} C^{\dag}$ with $C = -i e^{i \pi j_y} \tau_y$,
when the BdG wave function Eq.~\eqref{eq:luttingertsc-wfr} has the eigenvalue $E_{\ell i}$,
\begin{equation}
  C \Psi_{\ell i}^{\ast}({\vec r})
  = (2 \pi)^{-1/2} e^{-i (\ell + 3) \phi} [-v_{\ell i 3}, e^{i \phi} v_{\ell i 2}, -e^{2 i \phi} v_{\ell i 1}, e^{3 i \phi} v_{\ell i 0},
  u_{\ell i 3}, -e^{i \phi} u_{\ell i 2}, e^{2 i \phi} u_{\ell i 1}, -e^{3 i \phi} u_{\ell i 0}]^{\rm T}, \label{eq:luttingertsc-wfrc}
\end{equation}
has the eigenvalue $-E_{\ell i}$.
Therefore we only have to solve the radial BdG equation Eq.~\eqref{eq:luttingertsc-bdg} for $\ell \geq -1$.
The total number of particles $N$, the total spin $J_z$, and the bulk orbital AM $L_z$ can be calculated by
\begin{subequations} \begin{align}
  N
  = & \sum_{\ell = -1}^{\ell_{\rm max}} \sum_i \int_0^R r {\rm d} r
  [{\bm u}_{\ell i}^2(r) f(E_{\ell i}) + {\bm v}_{\ell i}^2(r) f(-E_{\ell i})], \label{eq:luttingertsc-nr} \\
  J_z
  = & \sum_{\ell = -1}^{\ell_{\rm max}} \sum_i \int_0^R r {\rm d} r
  [{\bm u}_{\ell i}^{\rm T}(r) j_z {\bm u}_{\ell i}(r) f(E_{\ell i})
  - {\bm v}_{\ell i}^{\rm T}(r) j_z {\bm v}_{\ell i}(r) f(-E_{\ell i})], \label{eq:luttingertsc-jzr} \\
  L_z
  = & \sum_{\ell = -1}^{\ell_{\rm max}} (\ell + 3/2) \sum_i \int_0^R r {\rm d} r [{\bm u}_{\ell i}^2(r) f(E_{\ell i}) - {\bm v}_{\ell i}^2(r) f(-E_{\ell i})] - J_z, \label{eq:luttinegrtsc-lzr}
\end{align} \label{eq:luttingertsc-r} \end{subequations}
\end{widetext}
respectively, in which $f(z) = \theta(-z)$ is the Fermi-Dirac distribution function at $T = 0$.
We use the central difference method with $N_{\rm g}$ grid points,
which reduces a set of the differential equations Eq.~\eqref{eq:luttingertsc-bdg} to the $8 N_{\rm g} \times 8 N_{\rm g}$ eigenvalue problem.
We put the disk radius $R = 50$, the number of grid points $N_{\rm g} = 500$, and the AM cutoff $\ell_{\rm max} = 512 - 2$.
Figure~\ref{fig:luttingertsc-eigen} shows the low-energy spectra for $C = 0, -3, -4$.
There are one PHS edge mode and one pair of the non-PHS edge modes in Fig.~\ref{fig:luttingertsc-eigen}(b), which results from $C = -3$.
On the other hand, there are two pairs of the non-PHS edge modes in Fig.~\ref{fig:luttingertsc-eigen}(c), which results from $C = -4$.

\section{Results} \label{sec:results}
\begin{figure*}
  \centering
  \includegraphics[clip,width=0.67\textwidth]{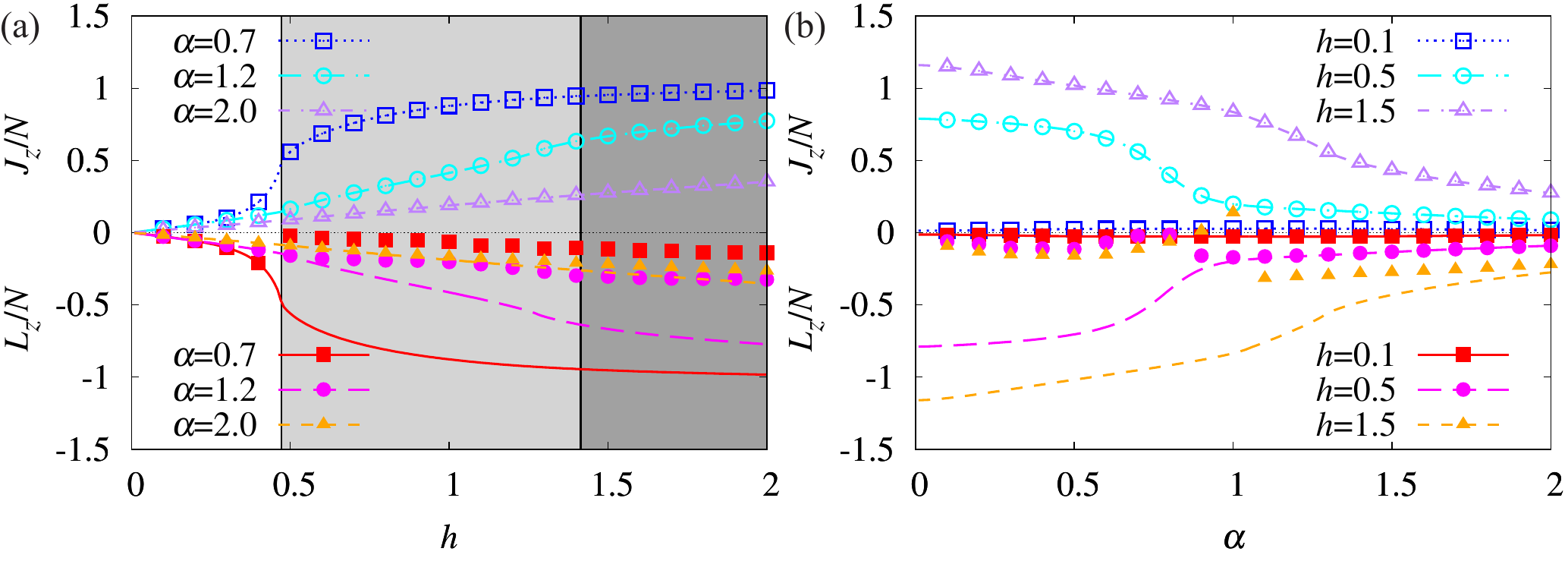}
  \caption{%
  (Color online) (a) $h$ and (b) $\alpha$ dependences of the total spin $J_z/N$ and the bulk orbital AM $L_z/N$.
  Symbols and lines are obtained by the real and reciprocal spaces, respectively.
  In (a), the light and dark gray regions indicate the topological phases with $C = -3, -4$, respectively.%
  } \label{fig:luttingertsc-am}
\end{figure*}

Figure~\ref{fig:luttingertsc-am} shows the $h$ and $\alpha$ dependences of the total spin $J_z/N$ and the bulk orbital AM $L_z/N$ obtained by the real and reciprocal spaces.
We find that $J_z/N$ obtained by the real space always coincides with that by the reciprocal space,
while $L_z/N$ obtained by the real space is reduced compared with that by the reciprocal space for $h > h_{\rm cH}$.
Therefore these two are denoted by $L_z^{(r)}/N$ and $L_z^{(k)}/N$, respectively, below.
This reduction of the bulk orbital AM is expected from the previous studies~\cite{PhysRevLett.114.195301,Volovik2014,PhysRevB.90.224519}
but not strong when $h$ is small or $\alpha$ is large.
This mechanism and quantitative criterion are discussed later.
We also find $L_z^{(k)}/N + J_z/N = 0$ as in an electron-doped topological SC~\cite{PhysRevB.92.024502}.

Owing to the presence of the SOIs, not the orbital AM $\ell_z$ but the total AM $i_z = \ell_z + j_z$ is conserved.
In the reciprocal space, we can easily check
\begin{align}
  [\ell_z, {\cal H}_{\vec k}]
  = & -[j_z, {\cal H}_{\vec k}]
  = 2 i [\sqrt{3} \xi_{2k} (\Gamma_3 \cos 2 \phi - \Gamma_4 \sin 2 \phi) \notag \\
  & + \alpha k (j_x \cos \phi + j_y \sin \phi)] \tau_z, \label{eq:luttingertsc-comm}
\end{align}
and $[i_z, {\cal H}_{\vec k}] = 0$ for each ${\vec k}$.
Therefore the eigenstates are the simultaneous eigenstates of $i_z$.
By summing up the eigenvalues of $i_z$ over the negative-energy eigenstates, we obtain the bulk total AM $I_z^{(k)} = L_z^{(k)} + J_z = 0$.
Such discussion holds only when the negative-energy eigenstates are well defined, namely, the system is gapped.
In the real space, Eq.~\eqref{eq:luttingertsc-r} is rewritten as
\begin{align}
  L_z^{(r)} + J_z
  = & -\frac{1}{2} \sum_{\ell = -1}^{\ell_{\rm max}} (\ell + 3/2) \sum_i [f(-E_{\ell i}) - f(E_{\ell i})] \notag \\
  & + \frac{1}{2} \sum_{\ell = -1}^{\ell_{\rm max}} (\ell + 3/2) \sum_i \int_0^R r {\rm d} r [{\bm u}_{\ell i}^2(r) - {\bm v}_{\ell i}^2(r)] \notag \\
  = & -\frac{1}{2} \sum_{\ell = -1}^{\ell_{\rm max}} (\ell + 3/2) \eta_{\ell}, \label{eq:rashbatsc-lzr2}
\end{align}
by using the spectral asymmetry,
\begin{equation}
  \eta_{\ell}
  \equiv \sum_i [f(-E_{\ell i}) - f(E_{\ell i})]. \label{eq:spec-asym}
\end{equation}
This is similar to the previous results on chiral SCs~\cite{Volovik1995,PhysRevLett.114.195301,Volovik2014}
but is modified by the presence of the SOIs~\cite{PhysRevB.93.174505}.
For $C = -3$ as shown in Fig.~\ref{fig:luttingertsc-eigen}(b), the non-PHS edge states cross $E = 0$ at $\pm (\ell_{\rm F} + 3/2)/R$
with $\ell_{\rm F}$ being called the Fermi AM~\cite{PhysRevLett.114.195301},
which leads to the nonzero spectral asymmetry $\eta_{\ell} = -2$ for $0 < (\ell + 3/2)/R < (\ell_{\rm F} + 3/2)/R$
and reduces the bulk orbital AM by $\Delta L_z^{(r)} = (\ell_{\rm F} + 2)^2/2$.

\begin{figure*}
  \centering
  \includegraphics[clip,width=0.67\textwidth]{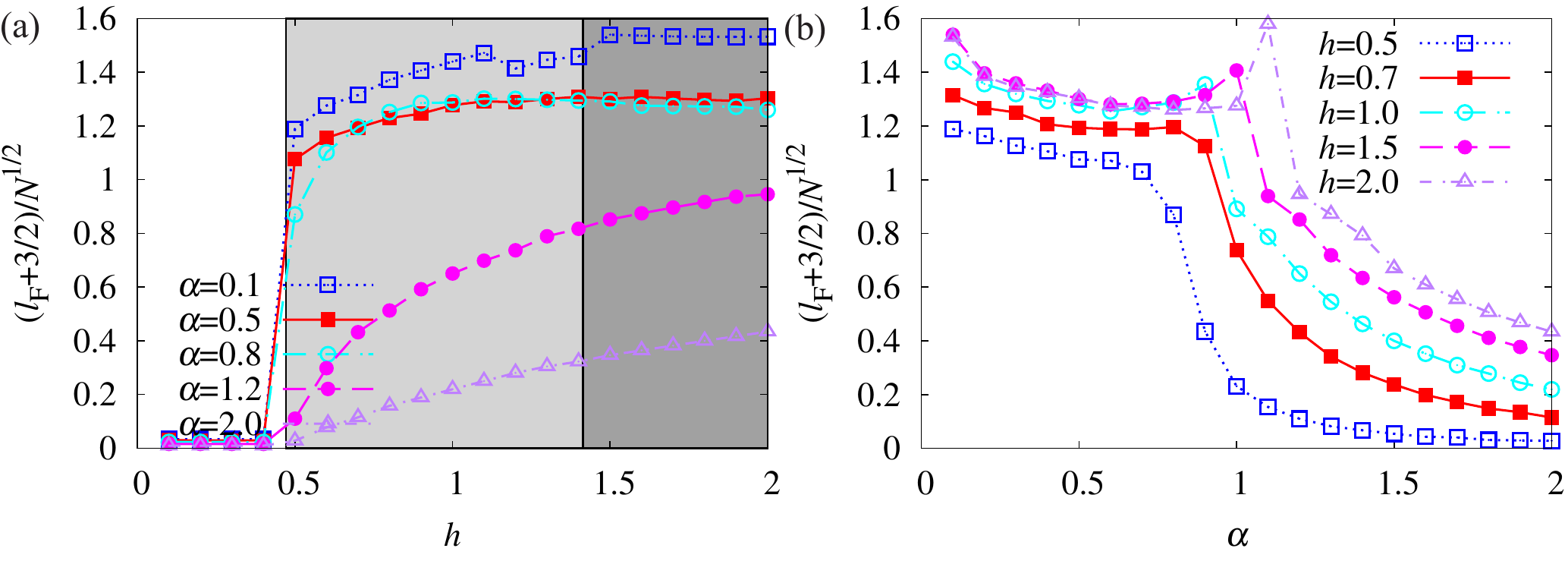}
  \caption{%
  (Color online) (a) $h$ and (b) $\alpha$ dependences of the Fermi AM.
  For $C = -4$, only the outer Fermi AM is shown because the inner Fermi AM is found to be small.
  In (a), the light and dark gray regions indicate the topological phases with $C = -3, -4$, respectively.%
  } \label{fig:luttingertsc-fermi}
\end{figure*}

\begin{figure*}
  \centering
  \includegraphics[clip,width=0.67\textwidth]{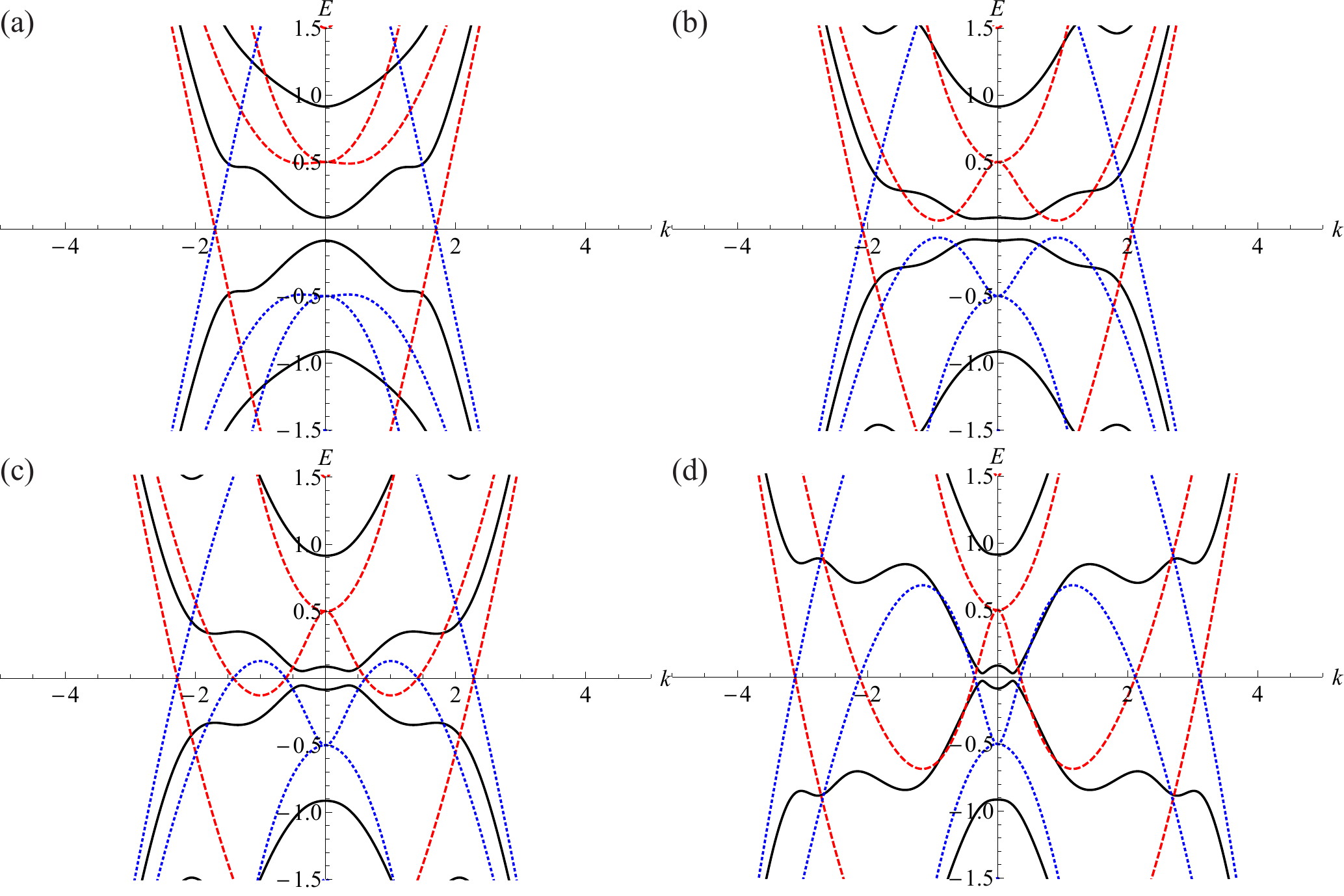}
  \caption{%
  (Color online) 
  Black solid line indicates the band structure for the BdG Hamiltonian Eq.~\eqref{eq:luttingertsc-hk},
  while red broken and blue dotted lines indicate the electron and hole bands in the normal state, respectively.
  We set $h = 0.5$ and
  (a) $\alpha = 0.5$,
  (b) $\alpha = 0.8$,
  (c) $\alpha = 0.9$,
  (d) $\alpha = 1.2$.
  Between (b) and (c), the Fermi AM shows the sharp decrease in Fig.~\ref{fig:luttingertsc-fermi}(b).%
  } \label{fig:luttingertsc-e0k}
\end{figure*}

Differently from chiral SCs, the reduction of the bulk orbital AM depends on $h$ and $\alpha$ through the Fermi AM.
To see this, we show the $h$ and $\alpha$ dependences of the Fermi AM normalized by $N^{1/2}$ in Fig.~\ref{fig:luttingertsc-fermi}
since we are interested not in $\Delta L_z^{(r)}$ but in $\Delta L_z^{(r)}/N$.
We find that the Fermi AM is large enough to reduce the bulk orbital AM almost to zero
for $(\lambda_{\rm L}^2 + h^2)/2 \lambda_{\rm L} < W - \mu$
but is small in the opposite case, leading to $L_z^{(k)}/N \simeq L_z^{(r)}/N$.
Here $\lambda_{\rm L} \equiv m (2 \alpha/\sqrt{3})^2/(\gamma_1 + 2 \gamma_2)$ is the effective strength of the Rashba SOI for the light-hole bands.
The factor $2/\sqrt{3}$ comes from the matrix elements in $j_x$ and $j_y$.
This is because the sharp decrease of the Fermi AM around $\alpha = 1$ is accompanied by the Lifshitz transition in the normal state.
The lowest particle band is always one of the heavy-hole bands with $j_z = +3/2$,
but the second lowest particle band changes from the other heavy-hole band with $j_z = -3/2$ to one of the light-hole bands with $j_z = +1/2$ at $h = W/2$.
In order to realize the topological phase with $C = -3$, $k = 0$ of the second lowest particle band should be above from the chemical potential.
However, when the strength of the Rashba SOI overcomes the Zeeman interaction, this light-hole band has its minimum at $k \not= 0$,
and hence the Lifshitz transition can occur as shown in Fig.~\ref{fig:luttingertsc-e0k}.
The non-PHS edge states connect the heavy-hole band with the antiparticle counterpart of the light-hole band before the Lifshitz transition.
Once the Lifshitz transition occurs, the non-PHS edge states connect the heavy-hole band with the inner part of the antiparticle counterpart of the light-hole band,
which leads to the sharp decrease of the Fermi AM.

The $h$ and $\alpha$ dependences of the bulk orbital AM in the reciprocal space $L_z^{(k)}/N$ in a hole-doped topological SC
can be understood in the same way as those in an electron-doped topological SC~\cite{PhysRevB.92.024502}.
We translate the previous results for electrons to heavy holes.
To do this, we define the effective strength of the Rashba SOI for heavy holes by $\lambda_{\rm H} \equiv m \alpha^2/(\gamma_1 - 2 \gamma_2)$.
For the small SOIs $2 \lambda_{\rm H} (\mu + W) + (3 h)^2 < \Delta_{\rm H}^2$ in the trivial phase,
the $s$-wave pairing potential is dominant, and hence we obtain $L_z^{(k)}/N \simeq 0$.
For $\lambda_{\rm H} < 3 h$ in the topological phase with $C = -3$, we expect $L_z^{(k)}/N \simeq -3/2$ from the $(f - i f)$- wave pairing potential.
In between, $L_z^{(k)}/N$ rapidly increases even in the trivial phase
since the $s$-wave pairing potential is no longer dominant due to the Pauli depairing effect, and the $(f \mp i f)$-wave pairing potentials coexist.
For the large SOIs $\lambda_{\rm H} > 3 h$, spins are tilted in the $xy$ plane, and $L_z^{(k)}/N$ is suppressed through the conservation law $L_z^{(k)} + J_z = 0$.
Note that $L_z^{(k)}/N$ shown in Fig.~\ref{fig:luttingertsc-am}(b) does not reach $-3/2$ even for $\lambda_{\rm H} < 3 h$
because the Zeeman interaction does not overcome the other SOI $\gamma_2 = 0.3$.

To summarize these results, the $h$ dependence of the bulk orbital AM is intriguing for the small SOIs as shown in Fig.~\ref{fig:luttingertsc-am}(a).
First, the system is trivial due to the $s$-wave pairing potential, and $L_z^{(r)}/N = L_z^{(k)}/N$ is almost zero.
Second, the system is still trivial due to the coexistence of the $(f \mp i f)$-wave pairing potentials, and $L_z^{(r)}/N = L_z^{(k)}/N$ rapidly increases.
Finally, the system turns into the topological phases with $C = -3, -4$, and only $L_z^{(r)}/N$ is strongly reduced by the non-PHS edge states.
Thus the $h$ dependence of $L_z^{(r)}/N$ in a hole-doped topological SC is nonmonotonic in contrast to that in an electron-doped topological SC.
On the other hand, for $2 \lambda_{\rm H} (\mu + W) + (3 h)^2 > \Delta_{\rm H}^2$, the system starts from the coexisting region.
In the topological phases, the reduction of $L_z^{(r)}/N$ does not work due to the Lifshitz transition in the second lowest particle band and the sharp decrease of the Fermi AM.
Therefore the $h$ dependence of $L_z^{(r)}/N \simeq L_z^{(k)}/N$ is almost monotonic and gradual
since $L_z^{(k)}/N = -J_z/N$ is suppressed by the large SOIs $\lambda_{\rm H} > 3 h$.

\section{Discussion} \label{sec:discuss}
So far, we have concentrated on the bulk orbital AM altogether.
Here we discuss how the PHS edge, non-PHS edge, and continuum states contribute to the bulk orbital AM separately.
The first and second lowest positive-energy eigenstates are continuum states in the trivial phase
and become one of the non-PHS edge states and the PHS edge state, respectively, in the topological phase with $C = -3$.
We can reveal the effects of the PHS and non-PHS edge states more clearly
by decomposing the bulk orbital AM with respect to the index of eigenstates around the topological phase transition.
Equation~\eqref{eq:luttinegrtsc-lzr} is decomposed as
\begin{align}
  L_z^{(r)}
  = & \int_0^R r {\rm d} r \ell_z^{\rm (r)}(r) \notag \\
  = & \int_0^R r {\rm d} r [\ell_z^{i = 2}(r) + \ell_z^{i = 1}(r) + \ell_z^{i \not= 1, 2}(r)]. \label{eq:luttingertsc-lzr2}
\end{align}
Remember that we have $8 N_{\rm g}$ eigenstates labeled by $i = -4 N_{\rm g} + 1, \dots, 0, 1, 2, \dots, 4 N_{\rm g}$,
$\ell_z^{i = 2}(r), \ell_z^{i = 1}(r), \ell_z^{i \not= 1,2}(r)$ are dubbed the PHS edge, non-PHS edge, and continuum contributions, respectively,
in the case of $C = -3$, as shown in Fig.~\ref{fig:luttingertsc-eigen}(b).
In the case of $C = 0$ as shown in Fig.~\ref{fig:luttingertsc-eigen}(a), all of them originate from the continuum states.

In the context of chiral $p$-wave SCs, Tsutsumi and Machida previously showed that the bulk orbital AM $L_z/N = -1/2$
consists of $L_z^{\rm MJ}/N = -1$ from the PHS edge state and $L_z^{\rm cont}/N = 1/2$ from the continuum states
and pointed out that these contributions can be experimentally distinguished by controlling the edge roughness~\cite{PhysRevB.85.100506}.
Thus the PHS edge state is expected to play an important role for the bulk orbital AM.
As shown in Ref.~\onlinecite{PhysRevB.92.024502} and Fig.~\ref{fig:luttingertsc-am}(a),
the bulk orbital AM increases in the trivial phase and does not jump at the topological phase transition apart from the reduction by the non-PHS edge states,
which indicates that the continuum contribution may change its sign from negative to positive at the topological phase transition.

\begin{figure*}
  \centering
  \includegraphics[clip,width=0.99\textwidth]{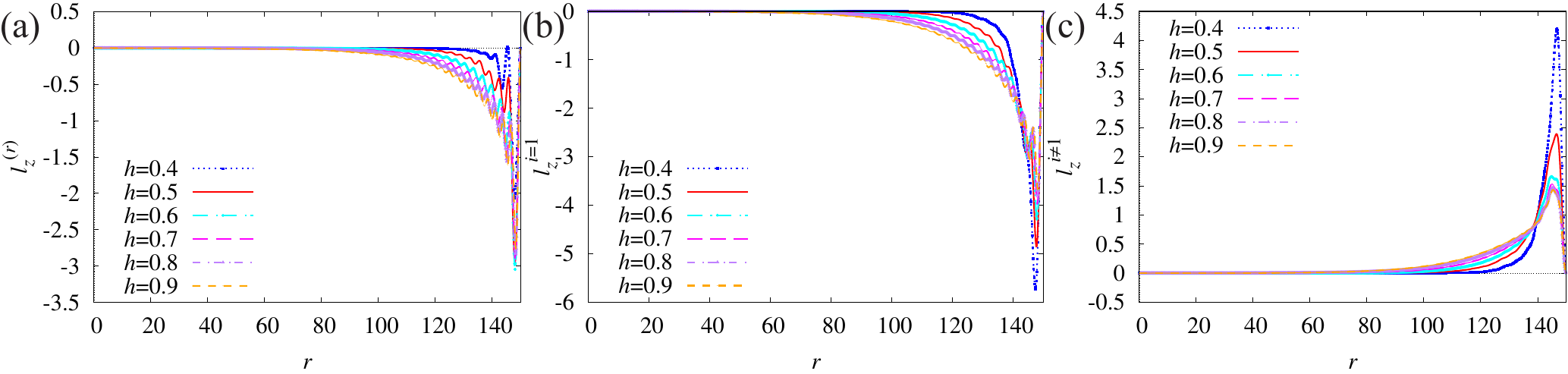}
  \caption{%
  (Color online) Radial distributions of
  (a) $\ell_z^{\rm (r)}(r)$,
  (b) $\ell_z^{i = 1}(r)$,
  and (c) $\ell_z^{i \not= 1}(r)$
  for $h < h_{\rm c} = 0.61$ ($C = 0$) and $h > h_{\rm c}$ ($C = -1$) in an electron-doped topological SC.
  We set $\alpha = 0.1$.
  See Ref.~\onlinecite{PhysRevB.92.024502} for the other parameters.%
  } \label{fig:rashbatsc-radial}
\end{figure*}

First, we carry out a similar analysis for an electron-doped topological SC,
which is designed to be a chiral $p$-wave SC with use of the Rashba SOI~\cite{PhysRevLett.101.160401,PhysRevLett.103.020401,PhysRevLett.104.040502},
in order to compare with a chiral $p$-wave SC and a hole-doped topological SC of our interest.
We set $\alpha = 0.1$ to focus on its chiral $p$-wave behavior and change $h$ around the phase transition point $h_{\rm c} = 0.61$.
See Ref.~\onlinecite{PhysRevB.92.024502} for the Hamiltonian and the other parameters we set.
In Fig.~\ref{fig:rashbatsc-radial}, we show the radial distributions $\ell_z^{i = 1}(r)$ and $\ell_z^{i \not= 1}(r)$.
For $C = -1$, the PHS edge contribution $\ell_z^{i = 1}(r)$ in Fig.~\ref{fig:rashbatsc-radial}(b)
and the continuum contribution $\ell_z^{i \not= 1}(r)$ in Fig.~\ref{fig:rashbatsc-radial}(c) are negative and positive, respectively,
which is consistent with the previous result obtained by the quasiclassical Eilenberger theory~\cite{PhysRevB.85.100506}.
However, we cannot find any significant difference between before and after the topological phase transition.
The $i = 1$ eigenstate, which goes down toward $E = 0$ and turns into the PHS edge state for $h > h_{\rm c}$,
is already localized at the edge even for $h < h_{\rm c}$ and has the negative orbital AM.
Therefore, as $L_z^{\rm MJ}/N$ from the PHS edge state in a chiral $p$-wave SC is expected to be fragile against the edge roughness~\cite{PhysRevB.85.100506},
not only the orbital AM from the PHS edge state in the topological phase but also that from one of the continuum states in the trivial phase is expected to be fragile.

\begin{figure*}
  \centering
  \includegraphics[clip,width=0.67\textwidth]{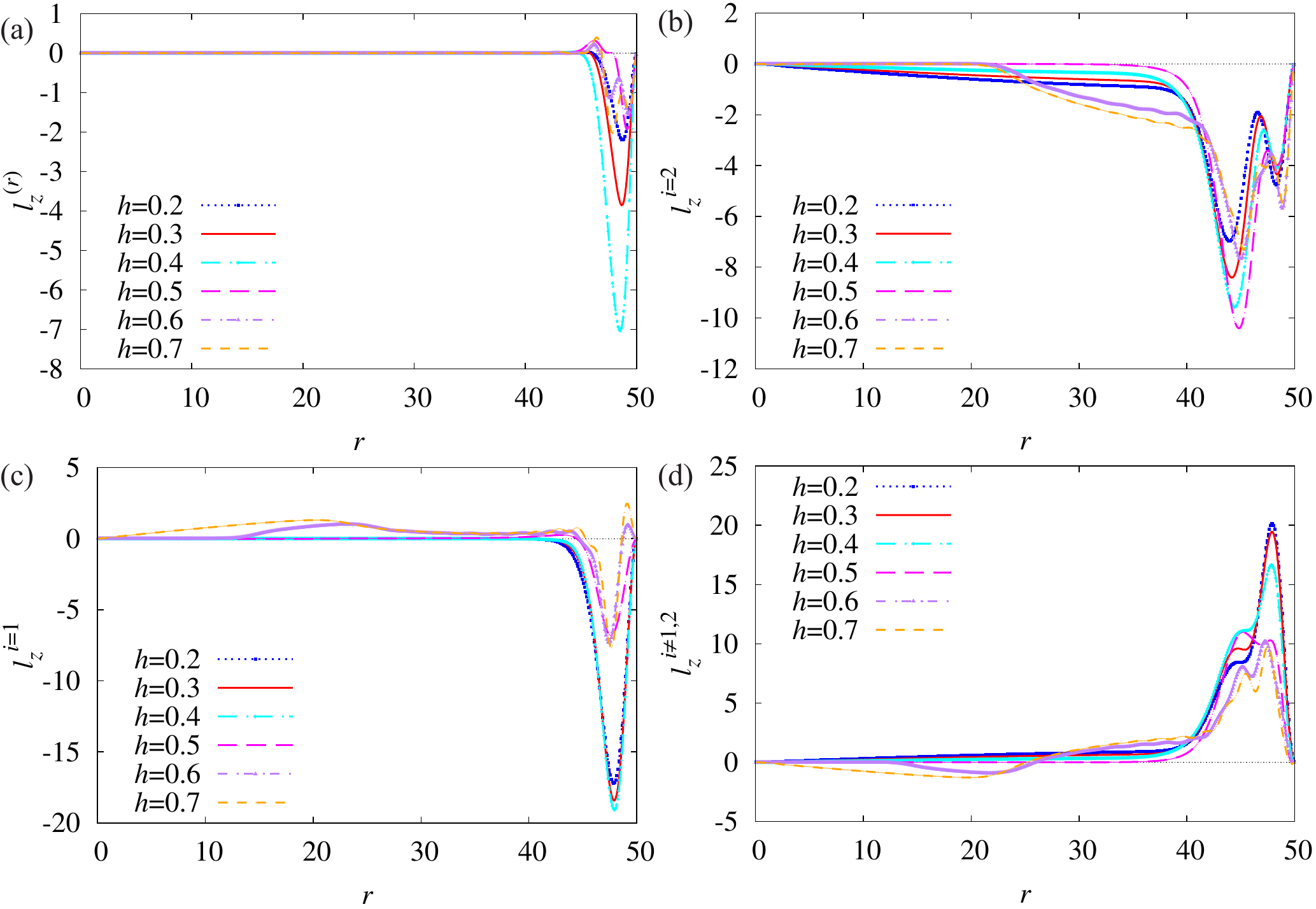}
  \caption{%
  (Color online) Radial distributions of
  (a) $\ell_z^{\rm (r)}(r)$,
  (b) $\ell_z^{i = 2}(r)$,
  (c) $\ell_z^{i = 1}(r)$,
  and (d) $\ell_z^{i \not= 1, 2}(r)$
  for $h < h_{\rm cH} = 0.47$ ($C = 0$) and $h > h_{\rm cH}$ ($C = -3$) in a hole-doped topological SC.
  We set $\alpha = 0.7$.
  } \label{fig:luttingertsc-radial}
\end{figure*}

Now let us move on to a hole-doped topological SC.
Figure~\ref{fig:luttingertsc-radial} shows the radial distributions $\ell_z^{i = 2}(r), \ell_z^{i = 1}(r), \ell_z^{i \not= 1,2}(r)$.
We set $\alpha = 0.7$ to avoid the accidental gap closing at $k \not= 0$ and to focus on its chiral $f$-wave behavior.
In Fig.~\ref{fig:luttingertsc-radial}(b), the orbital AM of the $i = 2$ eigenstate, which turns into the PHS edge state for $h > h_{\rm cH}$,
shows no significant difference between before and after the topological phase transition as in an electron-doped topological SC.
On the other hand, the radial distribution from the $i = 1$ eigenstate, which goes across $E = 0$ and turns into the non-PHS edge state for $h > h_{\rm cH}$,
discontinuously decreases at the topological phase transition.
This decrease is associated with the discontinuous reduction of the bulk orbital AM $\Delta L_z > 0$.
Note that the radial distribution is not positive because this eigenstate has the negative orbital AM even in the trivial phase similar to the $i = 2$ eigenstate.
Thus the effect of the non-PHS edge states can be found not in the radial distribution itself but in its change at the topological phase transition.

\section{Summary} \label{sec:summary}
In conclusion, we have investigated the bulk orbital AM in a two-dimensional topological SC
which is composed of a hole-doped semiconductor, a ferromagnet, and an $s$-wave SC and is characterized by $C = -3$.
We have obtained $L_z^{(k)}/N$ by using the Green's function formula of the Hall viscosity in the reciprocal space
and $L_z^{(r)}/N$ by solving the BdG equation on a circular disk.

Our findings are as follows:
(1) Before the topological phase transition, $L_z^{(r)}/N = L_z^{(k)}/N$ starts to increase as in an electron-doped topological SC~\cite{PhysRevB.92.024502}.
(2) For $\lambda_{\rm L} < W - \mu$,
$L_z^{(r)}/N$ is reduced from $L_z^{(k)}/N$ almost to zero owing to the non-PHS edge states with the large Fermi AM,
which is expected from the recent studies on chiral SCs with $\ell > 1$~\cite{PhysRevLett.114.195301,Volovik2014,PhysRevB.90.224519}.
As a consequence, $L_z^{(r)}/N$ is nonmonotonic as a function of the Zeeman interaction.
(3) In the opposite case, the Fermi AM of the non-PHS edge states becomes small associated with the Lifshitz transition in the normal state,
and the reduction of the bulk orbital AM does not work.
(4) This effect competes with the other effect to tilt spins and suppress the bulk orbital AM.
Thus a hole-doped topological SC offers an experimentally feasible stage
to investigate the bulk orbital AM in chiral $f$-wave SCs and the interesting deviations due to the SOIs.

Finally, let us comment on other models of two-dimensional topological SCs previously proposed.
SOIs are the most important ingredients throughout this paper, which make the system topological and modify the conserved quantity.
The models without use of SOIs~\cite{PhysRevB.88.184513,PhysRevB.90.064507,PhysRevB.93.035134} are out of our scope.
The explicit form of the conserved quantity depends on the details of the system as studied in Ref.~\onlinecite{PhysRevB.93.174505},
but it is given by $L_z + S_z$ at least for the models in Refs.~\onlinecite{PhysRevB.82.184516,PhysRevB.79.094504}.
Therefore we expect that our findings (1), (2), and (4) above are generic for these models with $C > 1$.
On the other hand, our finding (3) is quite specific to a hole-doped topological SC which makes use of $j = 3/2$ holes.
We also note that the Fermi AM and the reduction of the bulk orbital AM can be controlled by a confinement potential~\cite{PhysRevB.93.174505}.

\begin{acknowledgments}
  A.~S. was supported by the Japan Society for the Promotion of Science (JSPS) KAKENHI Grant No.~$25287085$,
  and Y.~N. was partially supported by JSPS KAKENHI Grant No.~$26800197$.
\end{acknowledgments}
%
\end{document}